\documentclass[twocolumn,prl,superscriptaddress,aps]{revtex4}
\usepackage{stmaryrd}
\usepackage{amsmath}
\usepackage{amssymb}
\usepackage{graphicx}
\usepackage{dcolumn}
\usepackage{bm}
\usepackage{hyperref}
\usepackage{color}

\begin{document}

\title{A large energy-gap oxide topological insulator based on the superconductor BaBiO$_3$}

\author{Binghai Yan}
\email{yan@cpfs.mpg.de}
\affiliation{Max Planck Institute for Chemical Physics of Solids, D-01187 Dresden, Germany}
\affiliation{Max Planck Institute for the Physics of Complex Systems, D-01187 Dresden, Germany}
\affiliation{Institute for Inorganic and Analytical Chemistry, Johannes Gutenberg University of Mainz, 55099 Mainz, Germany}
\author{Martin Jansen}
\affiliation{Max Planck Institute for Chemical Physics of Solids, D-01187 Dresden, Germany}
\author{Claudia Felser}
\affiliation{Max Planck Institute for Chemical Physics of Solids, D-01187 Dresden, Germany}

\begin{abstract}
{\bf Mixed-valent perovskite oxides based on
BaBiO$_3$~\cite{sleight1975,cava1988} (BBO) are, like cuperates,
well-known high-$T_c$ superconductors. Recent \textit{ab inito}
calculations~\cite{yin2011} have assigned the high-$T_c$
superconductivity to a correlation-enhanced electron--phonon
coupling mechanism, stimulating the prediction and synthesis of new
superconductor candidates among mixed-valent thallium
perovskites~\cite{yin2013,retuerto2013,schoop2013}. Existing
superconductivity has meant that research has mainly focused on
hole-doped compounds, leaving electron-doped compounds relatively
unexplored. Here we demonstrate through \textit{ab inito}
calculations that BBO emerges as a topological
insulator~\cite{qi2010,moore2010,hasan2010,qi2011RMP} (TI) in the
electron-doped region, where the spin-orbit coupling (SOC) effect is significant. BBO exhibits the largest topological energy gap of 0.7 eV among currently known TI materials~\cite{Yan2012rpp},  inside which Dirac-type topological surface states (TSSs) exit. As the first oxide TI, BBO is naturally stable against surface oxidization and degrading, different from chalcoginide TIs. An extra advantage of BBO lies in its ability to serve an interface between the TSSs and the superconductor for the realization of Majorana Fermions~\cite{fu2007c}.}
\end{abstract}

\maketitle

The parent compound BBO crystallizes in a mononclinic lattice that
is distorted from the perovskite structure, and this distortion is
attributed to the coexistence of two valence states, Bi$^{3+}$
($6s^2$) and Bi$^{5+}$ (6$s^0$), due to charge disproportion of
the formal Bi$^{4+}$. Octahedral BiO$_6$ breathes out and in for
Bi$^{3+}$ and Bi$^{5+}$, respectively~\cite{Cox1976}. Under
hole-doping conditions, such as in Ba$_{1-x}$K$_x$BiO$_3$ ($x \sim
0.4$)~\cite{cava1988,mattheiss1988} and BaBi$_{1-x}$Pb$_x$O$_3$ ($x
\sim 0.3$)~\cite{sleight1975,khan1977}, the breathing distortion is
suppressed, resulting in a simple perovskite lattice~\cite{Pei1990}
in which superconductivity emerges. The O-breathing phonon mode of
BiO$_6$ is believed to result in the pairing of superconducting
electrons in the Bardeen--Cooper--Schrieffer (BCS)
framework~\cite{yin2011}. In addition to the breathing distortion,
undoped BBO also presents extra O-tilting
distortions~\cite{Cox1976}, finally resulting in a mononclinic
phase. However, in ref.~\onlinecite{yin2011} and earlier
work~\cite{Mattheiss1983,Vielsack1996,Meregalli1998}, SOC was not
taken into account in the theoretical study, since the electronic
states in the superconducting (hole-doped) region mainly result from Bi-$6s$
and O-$2p$ orbitals whose SOC effect is usually negligible.

By including the SOC effect in density-functional theory
calculations of the BBO band structure, we discovered a band
inversion between the first (Bi-$6s$ state) and second (Bi-$6p$
state) conduction bands, which is stable against lattice
distortions. This inversion indicates that BBO is a
three-dimensional (3D) TI with a large indirect energy gap of 0.7 eV
when doped by electrons instead of holes. The band structure of
ideal cubic BBO (Fig. 1A) reveals that the conduction bands are
modified dramatically when SOC is included due to the presence of
the Bi-$6p$ states. The first conduction band crossing the Fermi
energy ($E_F$) has a considerable Bi-$6s$ contribution over the whole
Brillouin zone, except at the $R$ momentum point where the Bi-$6p$
contribution is dominant. Without SOC there is a zero energy gap at
$R$ because of the degeneracy of the $p$ states. When the SOC is
included, the $|p, j=3/2>$ and $|p, j=1/2>$ states split, which
results in the large indirect energy gap of 0.7 eV in the vicinity
of the $R$ point. Here, the
Bi-$6s$ state lies above the Bi-$6p$ states, causing band inversion
around the energy gap (Fig. 1A). Unlike bulk HgTe~\cite{bernevig2006d,koenig2007}, a well-known TI,
this inversion occurs between the $|s, j=1/2>$ state and the $|p,
j=1/2>$ state, rather than $|p, j=3/2>$ state. Since the Bi atom is
the inversion centre of the perovskite lattice, the $|s, j=1/2>$ and
$|p, j=1/2>$ states have ``+'' and ``$-$'' parities, respectively.
Thus, a TI state can be obtained if $E_F$ is shifted up into this
energy gap. The parities of all the valence bands below this gap
were also calculated at all time-reversal invariant momenta,
$\Gamma$ (0 0 0), $X$ (0.5 0 0), $M$ (0.5 0.5 0), and $R$ (0.5 0.5
0.5), which yielded $Z_2$ topological invariants (1;111), confirming
the topological nontrivial feature according to Fu and
Kane's~\cite{fu2007a} parity criteria. At a doping rate of one electron per formula unit, $E_F$ shifts inside the $s$--$p$ inversion gap, and
all the Bi$^{5+}$ ions become Bi$^{3+}$. Consequently, a cubic phase
appears when the BiO$_6$ breathing distortion is suppressed, similar
to the hole-doping case~\cite{Pei1990}. The new cubic lattice is found to
expand only slightly in comparison to the undoped lattice because
the Bi$^{3+}$--O bond is longer than the Bi$^{5+}$--O bond due to
the localized Bi-$6s$ orbital, while the band inversion is unaffected.

\begin{figure}
    \begin{center}
      \includegraphics[width=3.5 in]{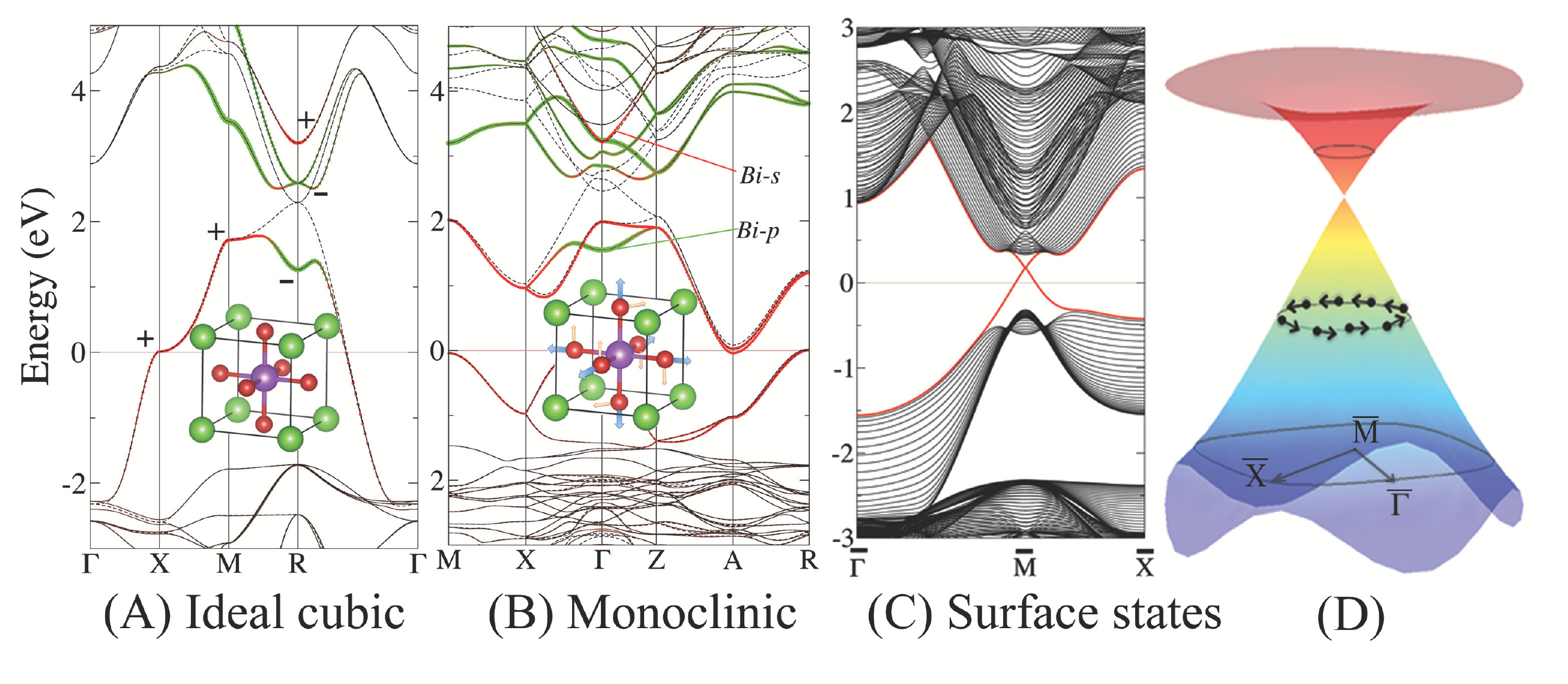}
    \end{center}
\caption{Band structures of bulk BiBaO$_3$ of the (A) ideal cubic
and (B) mononclinic structures. The solid and dashed lines represent
results with and without SOC, respectively, while the red and green
dots indicate the Bi-$s$ and Bi-$p$ states, respectively. The
parities are labelled in A. The Fermi energy is set to zero. The
lattice structures are shown in the insets, where Bi is shown in
purple, O in red, and Ba in green. The mononclinic structure
includes four formula units of BBO with O-breathing (blue arrows)
and O-tilting (yellow arrows) distortions from a cubic lattice.  (C)
The surface states of an electron-doped BiBaO$_3$ surface normalized
along the (001) direction. The red lines highlight the topological
surface states inside the bulk gap. (D) The surface Dirac cone near
$\bar{M}$. The Fermi surface is shown with a helical spin texture. }
    \label{fig:band}
\end{figure}

To illustrate the TSSs, we calculated the surface band
structure using a slab model. As an example, we take the surface to
be oriented along the (001) direction on which the bulk $R$ point is
projected onto the $\bar{M}$ point of the surface Brillouin zone.
The slab is thirty BBO units thick with the outermost atomic layers
being Ba-O. The TSSs, shown in Figs. 1C and D, exhibit a simple
Dirac-cone-like energy dispersion. The Dirac cone starts warping at
higher energies due to the cubic symmetry of the lattice. The Fermi
surface below the Dirac point exhibits a right-hand helical ``spin''-texture on the top surface, similar to that of
Bi$_2$Se$_3$-type TI materials~\cite{hsieh2009}, while the helicity corresponds to angular momenta $m_j=\pm 1/2$, instead of real electron spin. 
The Fermi velocity near the Dirac point is estimated to be approximately
$0.75\times 10^5$~m/s, and inside the large bulk energy gap, the TSSs
are well localized on the surface atomic layers to about two BBO
units or around 1 nm in thickness. On the other hand, to obtain the minimal effective model of the band topology, we
derive a four-band Hamiltonian similar to that for
Bi$_2$Se$_3$~\cite{zhang2009} in the basis of $|p; j=1/2, m_j=+1/2>, |s; j=1/2,
m_j=+1/2>,|p; j=1/2, m_j=-1/2>$, and $|s; j=1/2, m_j=-1/2>$:
\begin{equation}\label{eq:Heff3D}
   H({\bf k})=\epsilon_0({\bf k})\mathbb{I}_{4\times 4}+\nonumber
    \left(
    \begin{array}{cccc}
        \mathcal{M}({\bf k})&Ak_z&0&Ak_-\\
        Ak_z&-\mathcal{M}({\bf k})&Ak_-&0\\
        0&Ak_+&\mathcal{M}({\bf k})&-Ak_z\\
        Ak_+&0&-Ak_z&-\mathcal{M}({\bf k})
    \end{array}
    \right)        (1),
\end{equation}
where ${\bf k}={\bf k_0}-{\bf k_R}$ (0.5, 0.5, 0.5) is centred at
the $R$ point, and $k_\pm=k_x\pm ik_y$, $\epsilon_0({\bf
k})=C+Dk^2$, and $\mathcal{M}({\bf k})=M-Bk^2$. The main difference
from that of the Bi$_2$Se$_3$ Hamiltonian is that
Eq.~\ref{eq:Heff3D} is isotropic to ${\bf k}$ due to the cubic
symmetry. We obtain the parameters of Eq.~\ref{eq:Heff3D} by fitting the
energy spectrum of the effective Hamiltonian to that of the {\it ab
initio} calculations for the electron-doped cubic BBO using
$M=-0.625$ eV, $A=2.5$ eV\AA, $B=-9.0$ eV\AA$^2$, and $D=1.5$
eV\AA$^2$. Subsequently, the Fermi velocity of the TSSs is given by
$v=A/\hbar \simeq 0.5\times 10^5$~m/s, which is consistent with the
\textit{ab initio} calculations.

We can confirm that the TI phase is stable against lattice
distortions. The mononclinic BBO band structure (Fig. 1B) shows that
the $R$ point of the cubic lattice is projected onto the $\Gamma$
point due to band folding, but the $s$--$p$ band inversion at this
$\Gamma$ point is still present and the indirect gap is unchanged
(0.7 eV). Since the inversion strength (the energy difference
between the $|s, j=1/2>$ and $|p, j=1/2>$ bands) is nearly 2 eV (1.2 eV for the electron-doped structure),  the
TI phase is not destroyed by the O-breathing or -tilting distortions. Furthermore,
band structure calculations using the hybrid functional
method~\cite{Heyd2006,Krukau2006}, which is known to treat the dynamical
correlation effect well for BBO~\cite{Franchini2009,Franchini2010},
also validated the inversion for both the ideal cubic and distorted
structures. (Details are described in the Supplementary
Information.)

\begin{figure}
    \begin{center}
      \includegraphics[width=3in]{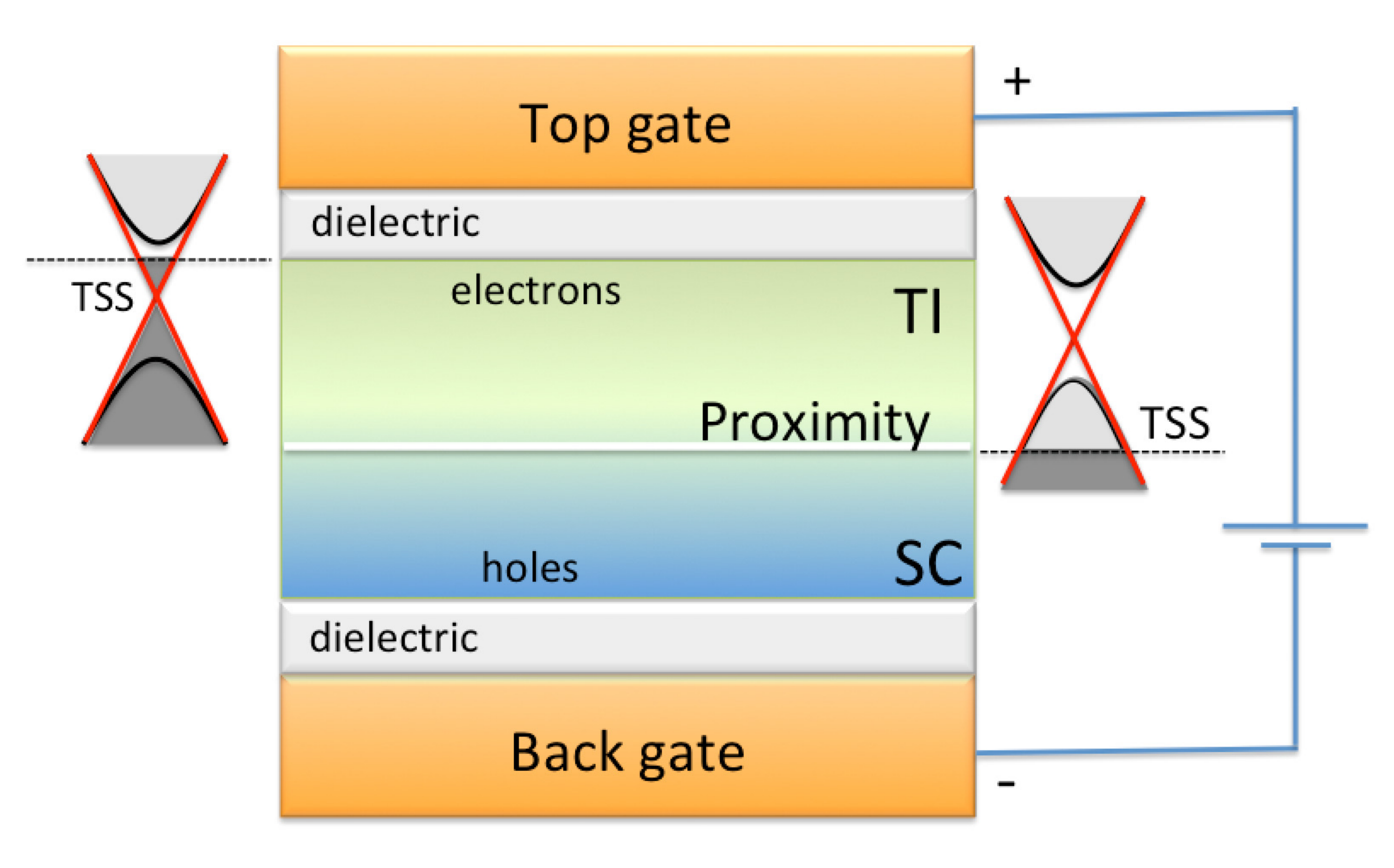}
    \end{center}
\caption{Schematic illustration of the interface between the topological
insulator (TI) and superconducting (SC) state in a gated thin
film device. The top and bottom surfaces are the TI and SC regions,
respectively. The position of the Fermi energy (dashed lines) shifts
down in the band structure. In the middle region, topological
surface states (TSSs) are interfaced with SC states and become SC
due to the proximity effect.}
    \label{fig:film}
\end{figure}

 Experimentally, electron-doped BBO
may be achieved in BaBi(O$_{0.67}$F$_{0.33}$)$_3$ by substituting O with F
atoms. The O and F atom have comparable atomic radii and electronegativityies, which can keep the octahedral BiO$_6$ stable.
For example, F subsitution for O was applied for iron-based superconductor LaOFeAs to realize electron-doping~\cite{Kamihara2008}.
 It is also possible to employ a state-of-art electrolyte gating technique to BBO to induce heavy electron-doping, which was realized for several mixed-valent compounds~\cite{Ye2009,Nakano2012,Jeong2013}.  In addition, oxygen vacancies, which are commonly observed in
experiments~\cite{Suzuki1985}, are also natural electron donors in
BaBiO$_{3-\delta}$. 
On the other hand, although the TSSs are unoccupied in pristine BBO
compounds, it may be possible to monitor these states directly via
monochromatic two-photon photoemission, as was recently employed to
monitor the empty TSSs of Bi chalcogenides~\cite{Niesner2012}.

Thus far we can state that BBO becomes a superconductor
with hole doping and a potential TI with electron doping. If
$pn$-junction-type devices are fabricated with BBO, an interface
between the TSSs and the superconductor may be realized, which is
necessary for the realization of Fu and Kane's~\cite{fu2007c}
proposal on Majorana Fermions for quantum computation. Here, we
outline a double-gated thin-film configuration, as illustrated in Fig.
2. If the bottom and top regions of the film are predoped as $p$ and $n$ type, respectively, 
the double-gated structure may feasibly induce a hole-rich bottom surface and an
electron-rich top surface, resulting in TSSs and superconductivity
states on the top and bottom surfaces, respectively. In the middle
region of the slab, the TSSs overlap with the bulk bands and
penetrate the bulk. These TSSs can then become superconducting as a
result of the proximity effect with the bottom superconducting
regime. Such a structure is likely to be attainable as  high-quality BBO thin films, which have been successfully grown on SrTiO$_3$~\cite{Sato1989,Gozar2007,Inumaru2008} and
MgO~\cite{Inumaru2008} substrates. And moreover, the O-tilting
lattice distortion was recently found to be suppressed in a BBO(001)
thin film on MgO~\cite{Inumaru2008}, which is very close to our
required cubic structure. 

The band structure of BaBiO$_3$ can act as a prototype for designing
new perovskite TIs. Ba can be substituted by Sc, Y, or La to obtain
new compounds as analogues of an electron-doped BBO. We found in
calculations that a similar band inversion exists in this case.
However, these compounds are semimetals (the Sc/Y/La-$d$ orbtials
are lower in energy than the Bi-$p$ states) and induce topological
semimetals. In contrast, CsTlCl$_3$-type halide perovskites, which
are predicted to be
superconductor candidates~\cite{yin2013,retuerto2013,schoop2013}, have band
structures that are similar to BBO. However, we did not observe
$s$--$p$ inversion for ATlX$_3$ (X = Cs, Rb, F, Cl, Br, or I),
because the SOC of Tl is not strong enough. When we can substitute Tl
with Sn or Pb,  we find that heavier members of this family,
such as CsSnI$_3$ and CsPbI$_3$, are near the boundary of a
topological trivial--nontrivial phase transition. Compressive
pressure is necessary to drive these boundary materials into the TI
region, which is consistent with recent theoretical calculations of
these halides ~\cite{Jin2012}.

 We thank Prof. X.-L. Qi at Stanford University and Prof. S. S. P. Parkin at IBM Almaden Research Center San Jose for fruitful discussions. B.Y. acknowledges financial support from the ERC Advanced Grant (291472) and computing time at HLRN Berlin/Hannover (Germany).


\appendix 
\section{Supplementary Information}

\section{Method}
In band structure calculations, we employed \textit{ab initio} density-functional theory (DFT) with the generalized gradient approximation (GGA)~\cite{perdew1996}. We employed the \textit{Vienna ab initio simulation package} with a plane wave basis~\cite{kresse1993}. The core electrons were represented by the projector-augmented-wave potential~\cite{kresse1999}.  For hybrid-funcitonal calculations, we adopted HSE06~\cite{Heyd2003,Heyd2006,Krukau2006} exchange-correlation functional and interpolated the band structures using Wannier functions~\cite{Mostofi2008}, where the DFT wavefunctions were projected to Bi-$sp$, Ba-$d$ and O-$p$ orbitals. The cubic crystal structure of BaBiO$_3$(BBO) was taken from ref.~\cite{Cox1979} with lattice constant $a=4.35$ \AA~ and the monoclinic structure was from ref.~\cite{Cox1976}. 

\section{Band structures of BBO with hybrid-funcntionals}

\begin{figure}
    \begin{center}
      \includegraphics[width=3.5 in]{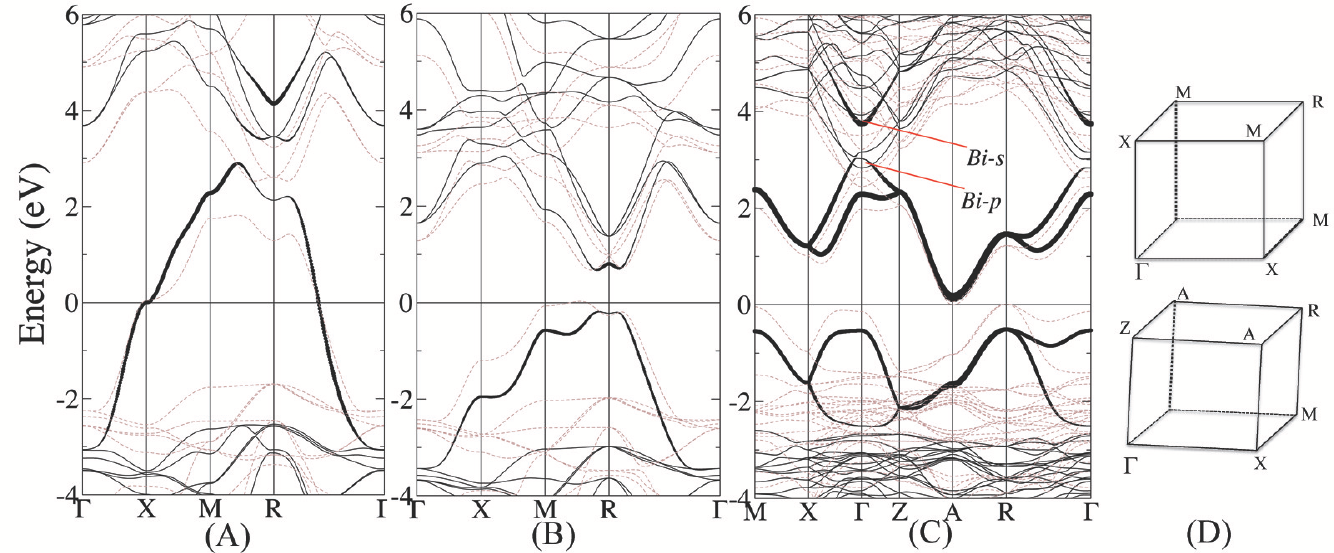}
    \end{center}
\caption{ Band structures of BBO calculated with HSE06 functionals for (A) ideal cubic, (B) electron-doped cubic (with lattice expansion), and (C) monoclinic structures. The thickness of black lines indicate the contribution of Bi-$s$ states. Results from GGA are also shown with braun dashed lines. SOC is only included for cubic lattices in A and B, but not for the monoclinic lattice in C.  (D) The first Brillouin zones are shown for the cubic (up) and monocinic (down) lattices. }
    \label{fig:film}
\end{figure}

We validated the band structures of both cubic and monoclinic lattices using HSE06 functionals.  For the cubic lattice, spin-oribt coupling (SOC) was also included. Compared to GGA, HSE06 was found to shift both the conduction and valence bands (see Figs. S1A and B), consist with ref.~\cite{yin2011}. One can see clearly that the Bi-$5s$ state is still above the energy gap, perserving the $s-p$ band inversion. From Fig.1 of the paper, SOC can dramatically modify the Bi-$6p$ bands, while it does not change the Bi-$s$ band. Therefore, the $s-p$ inversion will remain when SOC exists, if they are already inverted without SOC. Withoug SOC a clear feature of $s-p$ inversion is the $p$-state degeneracy at the $R$ point right above the Fermi energy (Fig. 1A of the paper). In many previous literatures that did not employ SOC, actually, this feauture can be clearly observed~\cite{Mattheiss1983,Takegahara1994,Vielsack1996,yin2011,Korotin2012}. So we performed HSE06 calculations without SOC for the monoclinic lattice, in order to reduce the compuational time of large supercells. Our result agrees with recent HSE06 calculations~\cite{Franchini2010}.
In Fig. S1C, one can also see the $s-p$ inversion. In all, we conclude that the band inversion is robust with hybrid-funcntionals for both ideal cubic and distorted lattices.

\section{Band structures of ABiO$_3$(A=Sc, Y and La)}
We calculated GGA band structures of ABiO$_3$(A=Sc, Y and La). 
The $s-p$ band inversion exists for all three compounds, in which the Fermi energy shifts into the inversion gap (See the band structures of ScBiO$_3$ and YBiO$_3$ for exmaples in Fig. S2). Therefore, they have the same nontrivial topological $Z_2$ invairants (1;1,1,1). However, the Sc/Y-$d$ states exhibit lower energy than Bi-$p$ states, resulting in a zero indirect gap. So these compounds are topological semimetals. In addition, we optimized the lattice constants of the perovskite lattice as 4.35, 4.41, and 4.46 \AA~ for Sc, Y and La compounds, respectively, before calculating band structures.  

\begin{figure}
    \begin{center}
      \includegraphics[width=3 in]{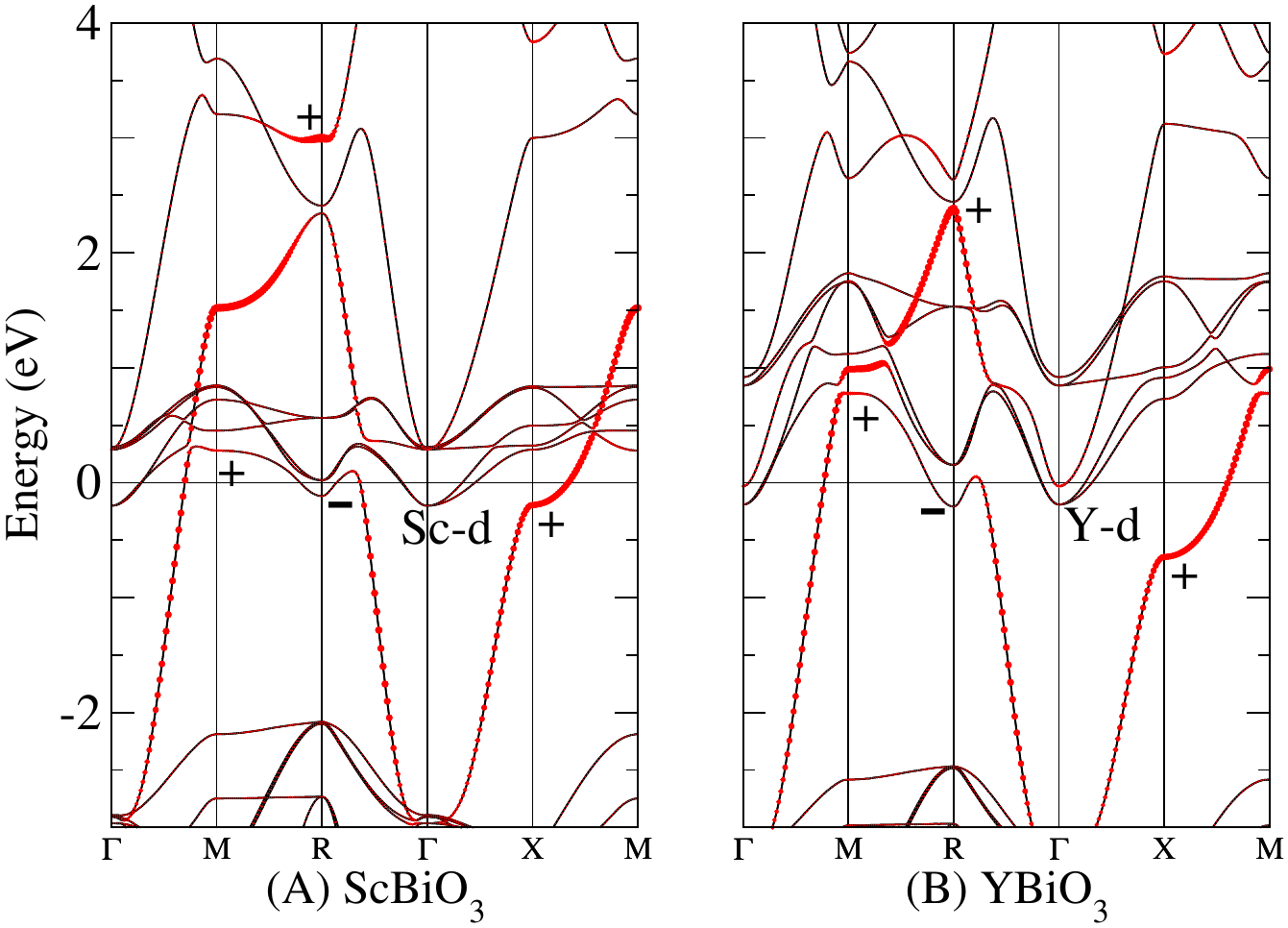}
    \end{center}
\caption{ Band structures of bulk ScBiO$_3$ and YBiO$_3$ calculated with GGA. Red balls present bands with Bi-s states. Parities are labeled. }
    \label{fig:film}
\end{figure}

\end{document}